\documentclass[sigconf,nonacm]{acmart}
\usepackage{setspace}
\AtBeginDocument{%
  \providecommand\BibTeX{{%
    \normalfont B\kern-0.5em{\scshape i\kern-0.25em b}\kern-0.8em\TeX}}}

\usepackage{balance} % For balanced columns on the last page
\newsavebox\tmpbox

\usepackage{caption}% or \usepackage{caption}
\usepackage{varwidth}
\renewcommand\footnotetextcopyrightpermission[1]{} % removes footnote with conference information in first column
\usepackage{booktabs} % For formal tables
\usepackage{amsmath}
\usepackage{float}
\usepackage[ruled,linesnumbered]{algorithm2e}

\setcopyright{none}

\begin{document}

\title{Improving Scalability of Contrast Pattern Mining for Network Traffic Using Closed Patterns}


%\author{Elaheh AlipourChavary\thanks{School of Computing and Information Systems, The University of Melbourne, Australia. ealipourchav@student.unimelb.edu.au, \{sarah.erfani, caleckie\} @unimelb.edu.au }
%\and Sarah M. Erfani\footnotemark[1]
%\and Christopher Leckie\footnotemark[1]}

\author{Elaheh AlipourChavary, Sarah M. Erfani, Christopher Leckie}
\email{ealipourchav@student.unimelb.edu.au, sarah.erfani@unimelb.edu.au, caleckie@unimelb.edu.au }
%\author{Sarah M. Erfani}
%\email{sarah.erfani@unimelb.edu.au}
%\author{Christopher Leckie}
%\email{caleckie@unimelb.edu.au}
\affiliation{%
 \institution{The University of Melbourne, Melbourne, Australia.}
 % \streetaddress{P.O. Box 1212}
 %\city{Melbourne}
  %\state{Victoria}
}

%\author{Lars Th{\o}rv{\"a}ld}
%\affiliation{%
 % \institution{The Th{\o}rv{\"a}ld Group}
%  \streetaddress{1 Th{\o}rv{\"a}ld Circle}
%\email{larst@affiliation.org}

%\affiliation{%
%  \institution{Inria Paris-Rocquencourt}

%\affiliation{%
% \institution{Rajiv Gandhi University}
 %\country{India}}

%\affiliation{%
%  \institution{Tsinghua University}
%  \streetaddress{30 Shuangqing Rd}
 % \country{China}}

%\affiliation{%
%  \institution{Palmer Research Laboratories}
%  \streetaddress{8600 Datapoint Drive}

%\affiliation{\institution{The Th{\o}rv{\"a}ld Group}}
%\email{jsmith@affiliation.org}

%\affiliation{\institution{The Kumquat Consortium}}
%\email{jpkumquat@consortium.net}


\begin{abstract}
  \textit{Contrast pattern mining} (CPM) aims to discover patterns whose support increases significantly from a \textit{background} dataset compared to a \textit{target} dataset. CPM is particularly useful for characterising changes in evolving systems, e.g., in network traffic analysis to detect unusual activity. While most existing techniques focus on extracting either the whole set of contrast patterns (CPs) or minimal sets, the problem of efficiently finding a relevant subset of CPs, especially in high dimensional datasets, is an open challenge. In this paper, we focus on extracting the \textit{most specific} set of CPs to discover significant changes between two datasets. Our approach to this problem uses \textit{closed patterns} to substantially reduce redundant patterns. Our experimental results on several real and emulated network traffic datasets demonstrate that our proposed unsupervised algorithm is up to 100 times faster than an existing approach for CPM on network traffic data \cite{chavary:summarizing}. In addition, as an application of CPs, we demonstrate that CPM is a highly effective method for detection of meaningful changes in network traffic.  
\end{abstract}

%\begin{CCSXML}
%<ccs2012>
%  <concept_id>10010520.10010553.10010562</concept_id>
%  <concept_desc>Computer systems organization~Embedded systems</concept_desc>
%  <concept_significance>500</concept_significance>
%  <concept_id>10010520.10010575.10010755</concept_id>
%  <concept_desc>Computer systems organization~Redundancy</concept_desc>
%  <concept_significance>300</concept_significance>
%  <concept_id>10010520.10010553.10010554</concept_id>
%  <concept_desc>Computer systems organization~Robotics</concept_desc>
%  <concept_significance>100</concept_significance>
%  <concept_id>10003033.10003083.10003095</concept_id>
%  <concept_desc>Networks~Network reliability</concept_desc>
%  <concept_significance>100</concept_significance>
%</ccs2012>
%\end{CCSXML}
%
%\ccsdesc[500]{Computer systems organization~Embedded systems}



\maketitle

\section{Introduction}

CPM (also is known as emerging pattern mining \cite{dong:efficient}) is an extension of \textit{frequent pattern mining} that extracts patterns whose support increases significantly from a background dataset to a target dataset (e.g., from day 1 to day 2) \cite{dong:efficient}. In other words, it searches for patterns that correspond to changes in the target dataset with respect to the baseline background dataset. CPM is useful in fields such as data summarization and network traffic analysis \cite{chavary:summarizing} to identify changes in a system. However, CPM is computationally expensive since (1) the \textit{Apriori property} does not hold for CPs, and (2) there are many candidate CPs in large datasets, especially for low support thresholds. Thus a major challenge is how to extract CPs in an efficient manner.

%The datasets under contrast can be different classes
%of a common dataset, or each dataset can have different classes. For example they
%can have two classes of normal or attack traffic. The datasets under contrast can
%also be the dataset of a single application that its data has been collected in different
%locations or different time periods.

Various techniques have been proposed in the literature for CPM, such as \cite{fan:fast, bailey:incremental, dong:efficient, seyfi:thesis}. However, the focus of most of these approaches is either extracting the most general patterns or all possible CPs, and they do not address the problem of redundancy and the computational cost of CPM. Alternatively, to extract a high-quality set of CPs and improve performance \cite{soulet:condensed, garriga:closed, li:DPMiner}, we can search for the most specific contrast patterns.
For example \textit{discriminative itemset mining} (which has a close relationship with CPM) uses the most specific patterns to extract discriminative itemsets. They use additional constraints such as productivity constraints and confidence-interval constraints to generate patterns \cite{kameya:towards, pham:statistically}.

In this paper, we focus on how to efficiently extract the most specific CPs to discover significant changes between two datasets (e.g., changes in network traffic across two different days). Our approach to this problem uses a compact representation of data, called \textit{closed patterns} \cite{pasquier:efficient}, which are those patterns that have no proper supersets with the same support. By elimination of minimal patterns in our approach, we considerably reduce the overlap between generated patterns, and by reducing the redundant patterns, we substantially improve the scalability of CPM. We propose a new scalable algorithm, called \textit{EPClose}, to extract CPs directly during closed pattern generation. We call this specific subset of contrast patterns as \textit{closed contrast patterns} (CCPs). In comparison with work in \cite{chavary:summarizing}, where CPs are generated by a post-mining process, we derive CPs directly during closed pattern generation. In particular, our aim is to examine whether closed patterns provide an expressive and efficient representation for CPM in practice. We
% We propose a new CCP mining algorithm called \textit{EPClose}, and we 
apply our {EPClose} algorithm to network traffic to investigate whether CCPs are useful for distinguishing attack traffic from normal traffic. 
 
Our experimental results show that although we are extracting the same set of CPs as the work in \cite{chavary:summarizing}, our proposed algorithm achieves a significant speed-up. In addition, our results demonstrate that CCPs have strong discriminative power in detecting pure patterns, i.e., most changes are either attack or normal traffic, but not a mixture of both. In summary, our main contributions are as follows:
\begin{itemize}
	\item We propose a new scalable algorithm, called \textit{EPClose} to extract the most specific contrast patterns (CCPs) directly from closed patterns.
	
	\item We show that CCPs are an expressive and efficient representation of CPs for network traffic analysis.

	\item We evaluate our algorithm and compare its performance with a baseline algorithm \cite{chavary:summarizing} on three network traffic datasets. The results demonstrate much better efficiency of our algorithm in comparison with \cite{chavary:summarizing}.
	
	\item We show the practicality of our algorithm in the application of network traffic summarization on different datasets. Although our CPM approach is unsupervised, our evaluation on several labeled datasets demonstrates the ability of CCPs to capture emerging attack patterns. The results show that derived CCPs are powerful tools for distinguishing the attack traffic from normal traffic.	
\end{itemize}

\section{Problem Statement}
Let $I=\{i_{1},i_{2}, \dotsc, i_{m}\}$ be the set of all distinct items in a \textit{dataset D}, where \textit{D} is a set of transactions and a \textit{transaction T} is a non-empty set of items. A transaction may occur several times in \textit{D}. An \textit{itemset} or \textit{pattern} \textit{X} is any subset of $I$. We use the terms itemset and pattern interchangeably throughout this work. An itemset \textit{X} is contained in a transaction \textit{T} if $X\subseteq{T}$. We define $D(X)=\{T \in D|X\subseteq T\}$.

The \textit{count} of \textit{X} in dataset \textit{D}, denoted as $count_{D}(X)$, is the number of transactions in dataset \textit{D} containing pattern \textit{X}. The \textit{support} of itemset \textit{X} is the fraction of transactions in dataset \textit{D} that contain \textit{X} and is given by $supp_{D}(X)=\frac{count_{D}(X)}{|D|}$. An itemset \textit{X} is frequent in a dataset \textit{D} if $supp_{D}(X)$ is greater than or equal to a pre-defined threshold $\sigma$. In the following definitions, let $supp_{i}(X)$ denotes $supp_{D_i}(X)$.  
\begin{definition}{
		The growth rate of a pattern \textit{X} for a target dataset $D_{t}$ compared to a background dataset $D_{b}$ is $gr(X,D_{t})= \frac{supp_{t}(X)}{supp_{b}(X)}$, where $gr(X,D_{t})=0$ if $supp_{t}(X)=supp_{b}(X)=0$, and $gr(X,D_{t})=\infty$ if $supp_{t}(X)>0$ and $supp_{b}(X)=0$.	
	}\label{def:growthRate}
\end{definition}
\begin{definition}{
		A contrast pattern \textit{X} is a pattern whose \textit{support} is significantly different from one dataset to another. Given a growth rate threshold $\rho>1$, pattern \textit{X} is a contrast pattern for dataset $D_{t}$ if $gr(X,D_{t})\,\geq \, \rho$.}\label{def:contrastPattern}
\end{definition}

For example, suppose we are given two datasets $D_b$ and $D_t$ shown in Table \ref{table:Example} with five transactions in each dataset. Each transaction is a subset of the itemset $I=\{a, b, c, d, e, f, g\}$. Also, suppose for all examples of this paper $\sigma=0.4$ and $\rho=1.5$. We are interested in CPs from the background dataset $D_{b}$ to the target dataset $D_{t}$. Hence, we need to extract all patterns $X$ whose $gr(X,D_{t}) \geq \rho$. For example, the patterns $\{c,e\}(3:1)$\footnote{\label{cpExample} Given $k\geq 1$ $\{a_{1},a_{2},...a_{k}\}(n:m)$ shows that the pattern $\{a_{1},a_{2},...a_{k}\}$ repeats $n$ times in $D_{t}$ and $m$ times in $D_{b}$} is a CP with $gr=3/1$ and $\{a,b,c,e\}(2:0)$ is another CP with $gr=\infty$.

%\begin{table}
%	\centering
%	\caption{Example datasets for contrast patterns}
%	\label{table:Example}	
%	\begin{tabular}{cccc}
%		\toprule
%		$D_p$ & $D_n$ \\ \midrule
%		abd& abf\\
%%		\hline
%		bce & bce  \\	
%%		\hline	
%		abce & bcfg \\	
%%		\hline	
%		be & bc  \\	
%%		\hline	
%		abce & abd  \\				
%		\bottomrule
%	\end{tabular}
%\end{table}
%
%\begin{figure}[ht]
%	\centering
%	%\includegraphics[width=\linewidth]{ProcessingTime}
%	\includegraphics[height=3.8cm,width=6cm]{FPTree}
%	%	\vspace*{-10pt}	
%	\caption{ An example of FP-Tree.\label{fig:FP-Tree}}
%	\Description{An example of FP-Tree.}
%	\vspace{-1.1em}	
%\end{figure}

\vspace{-1.0em}	
\begin{table}[ht]
	\begin{varwidth}[b]{0.2\linewidth}
		\centering
		\caption{Example datasets}
		\vspace{-1.0em}
		\begin{tabular}{ c c}
			\toprule
			$D_b$ & $D_t$ \\ \midrule
			abf&abd\\			
			bce&bce \\			
			bcfg&abce  \\				
			bc&be  \\				
			abd&abce   \\	
			\bottomrule
		\end{tabular}		
		\label{table:Example}
	\end{varwidth}%
	\hfill
	\begin{minipage}[b]{0.7\linewidth}
		\centering
		\includegraphics[height=3.8cm,width=6cm]{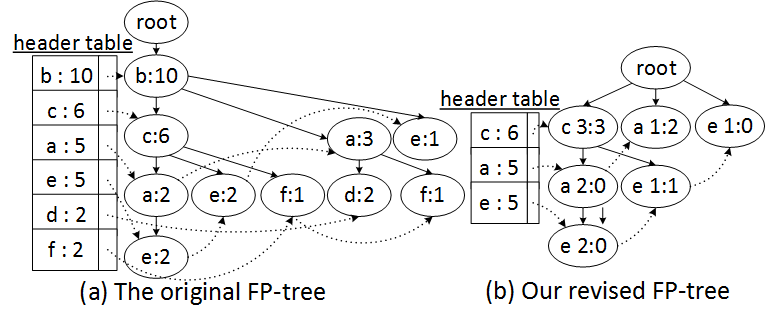}
		\vspace{-2.4em}
		\captionof{figure}{ An example of FP-tree}
		\label{fig:FPTree}
		\Description{An example of FP-tree.}
	\end{minipage}
\end{table}
\vspace{-0.8em}

For extracting CPs, our approach is to use \textit{closed patterns}, which are those patterns that have no proper supersets with the same support. For a formal definition, we utilize the \textit{closure operator} $\mathcal{H}$ such that $ \mathcal{H}(X,D)=\{ \bigcap T | T \in D(X) \}$. A pattern $X$ is closed if $X=\mathcal{H}(X,D)$, i.e., a pattern is closed if it is equal to its closure. 
%A closed pattern $x$ is called a \textit{Frequent Closed Itemset} (FCI) if $supp_{D}(X) \geq \sigma$ , that $\sigma$ is a predefined support threshold. 
%We introduce \textit{closed contrast patterns} (CCPs) which are the most specific CPs. 

\begin{definition}{
		Given two datasets $D_b$ and $D_t$, with size $n_b$ and $n_t$ respectively, the minimum support threshold of $\sigma$, and the growth rate threshold of $\rho>1$, a pattern \textit{X} is a CCP from $D_b$ to $D_t$ if it satisfies the following conditions:
		
		(1) $count_{D_t}(X)\geq \sigma n_t$;
		
		(2) $\mathcal{H}(X,D)=X 
		\;\;\;\;\;\;\;\;\;  
		where \; D=D_b \cup D_t$;
		
		(3) $gr(X,D_t)\,\geq \, \rho$

	}\label{def:CCP}
\end{definition}

The first condition guarantees to eliminate infrequent itemsets w.r.t. the target dataset. The second condition ensures that the pattern is closed and the last condition identifies only CPs.  

\textit{Problem statement}: Given two datasets of $D_b$ and $D_t$, the minimum support threshold of $\sigma$, and the growth rate threshold of $\rho>1$, how can we extract CCPs efficiently from $D_b$ to $D_t$. 

For example in Table \ref{table:Example}, the pattern $\{c,e\}(3:1)$ is a CP but it is not a CCP, since it does not satisfy condition 2 of Definition \ref{def:CCP}. The closure of this pattern is  $\{b,c,e\}(3:1)$. It implicitly conveys that the pattern $\{c,e\}$ will not appear in a transaction without $\{b\}$. Therefore, non-closed patterns are considered as redundant. However, the pattern $\{b,c,e\}(3:1)$ is a CCP with $gr=3/1$. Thus, our aim is to derive all CPs that are also closed.

\section{Our Approach: EPClose}
In this section, we investigate how to derive CCPs efficiently during closed pattern generation.

\subsection{Contrast Pattern Mining}\label{sub:CPM}
Having a pair of datasets $D_b$ and $D_t$, a naive method for extracting CPs from the closed patterns is to first discover all closed patterns of each dataset separately; then, as a post-processing step, match the two sets of closed patterns to find similar patterns in the two datasets and compute their supports; and finally compute the growth rate of similar patterns according to their support to find the collection of CPs \cite{chavary:summarizing}. However, in large and high-dimensional datasets the matching step is computationally expensive. 

To overcome this problem, we propose an algorithm \textit{EPClose}, which modifies a closed pattern mining algorithm called \textit{FP-close} \cite{grahne:fpclose}, such that we can extract CPs directly during closed pattern generation. \textit{FP-close} is a depth-first algorithm that uses \textit{FP-growth} \cite{han:mining} recursively to mine closed frequent itemsets (CFI). An itemset \textit{X} is a CFI in a dataset \textit{D} if it is closed and its support is greater than or equal to the minimum support threshold. It uses an efficient data structure called an \textit{FP-tree} to compress the dataset in memory. However, our revised version of \textit{FP-tree} has two differences with the original \textit{FP-tree}. The first difference is that the original \textit{FP-tree} keeps three fields in each node: \textit{item-name}, \textit{count} and \textit{node-link}. We replace the \textit{count} with the two counts of background and target datasets separately. The second difference is that the original \textit{FP-tree} is constructed from all frequent items, while we borrow the concept of full support items (FSIs) from \cite{pei:closet}, and remove FSIs from the \textit{FP-tree} construction. FSIs are those items that appear in each transaction of a dataset (implicitly they are frequent). However, unlike \cite{pei:closet}, we use it not only for \textit{conditional projected datasets}\cite{han:mining}, but also for the original dataset used in \textit{base FP-tree} construction. FSIs have the following property: 

\begin{property}\label{propertyFSI}
	The set of FSIs generates a candidate closed frequent itemset (CFI). If the newly discovered CFI is not a subset of any previously discovered CFI with the same \textit{count}, it is marked as a CFI. 
\end{property}

\textit{Join dataset}: The \textit{FP-close} algorithm derives closed patterns from a single dataset, whereas our objective is to compare two datasets and find the differences between them. Thus, we assume that we merge the target dataset $D_t$ and the background dataset $D_b$ into a single join dataset $D=D_b\cup D_t$. By considering the join dataset, all CCPs should not only be frequent in the join dataset, but also should be frequent in the target dataset (the first condition of Definition \ref{def:CCP}).

An example of the original \textit{FP-tree} and our modified \textit{FP-tree} is presented in Figure \ref{fig:FPTree}, which is constructed from Table \ref{table:Example}. Each node corresponds to one item. For each node, we also keep the item counts, separately, for two datasets (Figure \ref{fig:FPTree}(b)). It is clear from the figure that the size of our \textit{FP-tree} is smaller than the original one. The reason is that item $b$ is a FSI, and items $d$ and $f$ are infrequent in $D_t$, so we removed them from our \textit{FP-tree}. This early pruning can reduce the size of the \textit{FP-tree} considerably. Please refer to \cite{han:mining} for details of \textit{FP-tree} construction.

\subsubsection{EPClose Algorithm}
Before applying the recursive procedure of \textit{EPClose()}, the algorithm first scans the join dataset $D=D_b\cup D_t$ and counts the frequency of each item for $D_{b}$ and $D_{t}$ separately, and saves them in an \textit{F-list} according to their \textit{frequency in descending order}. Then, it finds all FSIs from the \textit{F-List} and according to Property \ref{propertyFSI}, marks the set of FSIs as a closed pattern and saves it in a \textit{CFI-tree}, which is a tree for saving closed patterns. Infrequent items in $D_{t}$ are also removed from the \textit{F-list}. These two early pruning steps considerably reduce the size of the \textit{FP-tree} in the \textit{EPClose} algorithm. The pseudo-code of \textit{EPClose} is shown in Algorithm \ref{alg:EPClose}. The method takes an \textit{FP-tree}, denoted as \textit{FPT}, as an input. \textit{FPT} has two attributes: $FPT.header=\{a_{1},a_{2}, \dotsc, a_{k}\}$ and \textit{FPT.base}. \textit{FPT.header} is the header table of the \textit{FP-tree}, and \textit{FPT.base} is an itemset for which \textit{FPT} is a conditional \textit{FP-tree}. In \textit{X's conditional FP-tree}, denoted as $FPT_{X}$, $base$ is equal to the pattern of $X$.

\begin{algorithm}[!ht]
	\KwData{FPT: FP-tree,
		C: CFI-tree, 
		FSI: list of full support items,
		$\sigma$: support threshold,
		$\rho$: gr threshold
	}
	\KwResult{\textit{CCP-List}: closed contrast patterns list		
	}
	
	\uIf{ FPT only has a single branch B}
	{
		generate all CFIs from B and FSIs, and call CCP-checking()\; \label{alg:closedChecking1}		
	}
	\textbf{else} \For{all item $a_{i} \in FPT.header=\{a_{1}, a_{2}, \dots, a_{k}\}$}
	{ 	
		set $\beta=FPT.base \cup a_{i}$ and 
		$betaCount\!=\!min(count_D(FPT.base), count_D(a_{i}))$\;		
		\If{closed-checking($\beta \cup FSI$,C)==fail}{
			construct $a_{i}$'s conditional pattern base, and
			count the frequency of items and save in $frequencyMap=\{i_{1}, i_{2}, \dots, i_{m}\}$\;\label{alg:countPhase}
			
			%create F-list:
			\For{each item $i_{j} \in frequencyMap$}
			{
				\uIf{$count_D(i_{j})==betaCount$}
				{
					insert $i_{j}$ to $FSI_{\beta}$\;
					remove $i_{j}$ from frequencyMap\;
				}		
				
				\label{alg:optimization}
				\textbf{else} \If{ $count_{D_{t}}(i_{j})\leq \sigma|D_t|$} 
				{
					remove $i_{j}$ from frequencyMap\;
				}	
			}
			copy frequencyMap in F-list\;
			$ Z=\beta \cup FSI_{\beta} \cup FSI$ \;
			\If{closed-checking(Z,C)==fail}{
				insert Z to C and call
				CCP-checking()\;\label{alg:EPChecking2}
			}
			
			construct $\beta$'s conditional FP-tree $FPT_{\beta}$\;
			\If{$FPT_{\beta} \neq \emptyset $}{
				$ FSI=FSI_{\beta} \cup FSI$\;
				call EPClose($FPT_{\beta}$, C, FSI, $\sigma$, $\rho$)\;
			}
		}
	}	
	\caption{EPClose Algorithm\label{alg:EPClose}}	
\end{algorithm}

\setlength{\textfloatsep}{1\baselineskip}

During the recursion, if \textit{FPT} has a single branch \textit{B}, the algorithm generates all CFIs from \textit{B} and the FSIs according to \textit{FP-close}, and then applies the \textit{CCP-Checking} function. This function examines the conditions of Definition \ref{def:CCP}; if pattern $X$ satisfies all conditions, then it is marked as a CCP and saved in the \textit{CCP-List} along with its corresponding $count_{D_{b}}(X)$ and $count_{D_{t}}(X)$. If \textit{FPT} is not a single branch, the algorithm is prepared for another recursive call by constructing $\beta$'s conditional FP-tree, denoted as $FPT_{\beta}$. Unlike \textit{FP-close}, the \textit{EPClose} algorithm calls the \textit{closed-checking} function before constructing $a_{i}$'s conditional pattern base, and if $\beta \cup FSI$ passes the closed checking, $a_{i}$'s conditional pattern base is constructed. In the \textit{closed-checking} function, if a pattern does not have any superset with the same support in the CFI-tree \textit{C}, it will be marked as a closed pattern \cite{grahne:fpclose}. 

\textit{EPClose} saves the dataset distribution information of existing items in the projected database as a hash map, denoted as \textit{frequencyMap}, in the form of $frequencyMap~=~<key,value>$ according to line \ref{alg:countPhase}, where $key$ is an item $i_j$ and $value$ is an array of two counts of $i_j$ in $D_{b}$ and $D_{t}$. After this, \textit{EPClose} finds local FSIs, and moves them from the \textit{frequencyMap} to a local $FSI_{\beta}$. Then the algorithm removes all infrequent items from $D_{t}$, according to line \ref{alg:optimization}. By using these two methods of pruning local FSIs and discarding infrequent items in $D_{t}$, the size of $FPT_{\beta}$ can be considerably reduced. Finally, before construction of $\beta$'s conditional FP-tree, the \textit{EPClose} algorithm executes an extra \textit{closed-checking} to determine if the new suffix pattern of $Z=\beta \cup FSI_{\beta} \cup FSI$ is a closed pattern or not. Then the algorithm constructs $\beta$'s conditional FP-tree and after merging the local $FSI_{\beta}$ and global FSI, calls the recursive method of \textit{EPClose} for $FPT_{\beta}$.

By applying algorithm \textit{EPCLose} to Table \ref{table:Example}, we derive 4 CCPs from all the patterns. They are $\{b,e\}(4:1),\{a,b\}(3:2),\{b,c,e\}(3:1),\{a,b,c,e\}(2:0)$. The patterns $\{b\}(5:5)$ and $\{b,c\}(3:3)$ are closed patterns, but they are not CCPs, since they do not satisfy condition 3 of Definition \ref{def:CCP}.

%\vspace{-1.5em}
\section{Experimental Results}
To evaluate the efficiency of the proposed \textit{EPClose} algorithm, we compare it with the \textit{ExtCP} algorithm \cite{chavary:summarizing}. For empirical evaluation, three benchmark network traffic datasets are used, namely Kyoto 2006+\footnote{\label{footnote1}http://www.takakura.com/Kyoto\_data/}, KDD-CUP 1999\footnote{http://kdd.ics.uci.edu/databases/kddcup99/kddcup99.html} and BGU\footnote{https://archive.ics.uci.edu/ml/datasets/detection\_of\_IoT\_\\botnet\_attacks\_N\_BaIoT}. Kyoto is a real dataset, while the other two are emulated. Table \ref{table:datasets} provides the parameters of each dataset. In the Kyoto dataset, we considered the traffic of 15 and 16 July 2007 as the background and target datasets, respectively. For BGU and KDD'99 we randomly select the target and background datasets. The continuous attributes of datasets were discretized by the \textit{equal-frequency unsupervised discretization} method, and the number of bins in discretization has been given in Table \ref{table:datasets}. The growth rate threshold is set to 5 for Kyoto and KDD'99 and 1.5 for BGU. All experiments were run in Java on a 2.6GHz CPU with 16GB of memory running Windows 7.

\begin{table}
	\centering
	\caption{Dataset information}
%	\vspace{-1.0em}
	\label{table:datasets}
	\begin{tabular}{lccccc}
		\toprule
		Dataset & $|D_b|$ & $|D_t|$ & Attributes & Bins & Items\\ \midrule
		Kyoto &119702&123835&14& 4&108\\			
		KDD'99 &15000&15000&10 & 5&38 \\
		BGU &4500&4500&23 & 2&46 \\
		\bottomrule
	\end{tabular}
\end{table}

%\begin{figure*}[ht]
%	\centering
%	%\includegraphics[width=\linewidth]{ProcessingTime}
%	\includegraphics[height=3.8cm,width=17cm]{ProcessingTime}
%%	\vspace*{-10pt}	
%	\caption{Processing Time of different datasets.\label{fig:ProcessingTime}}
%	\Description{Processing Time of different datasets.}
%	\vspace{-1.1em}	
%\end{figure*}
%
%\begin{figure*}[ht]
%	\centering
%	%\includegraphics[width=\linewidth]{ProcessingTime}
%	%	\includegraphics[height=3.8cm,width=17cm]{AttackRatio}
%	\includegraphics[height=3.8cm,width=\textwidth]{AttackRatio}
%	
%	%	\vspace*{-10pt}	
%	\caption{Attack Ratio for different datasets.\label{fig:AttackRatio}}
%	\Description{Attack Ratio for different datasets.}
%	\vspace{-1.1em}	
%\end{figure*}
%\vspace{-0.6em}
\begin{figure}[ht]
	\begin{minipage}[b]{0.47\linewidth}
		\centering
		\includegraphics[height=6.1cm,width=\textwidth]{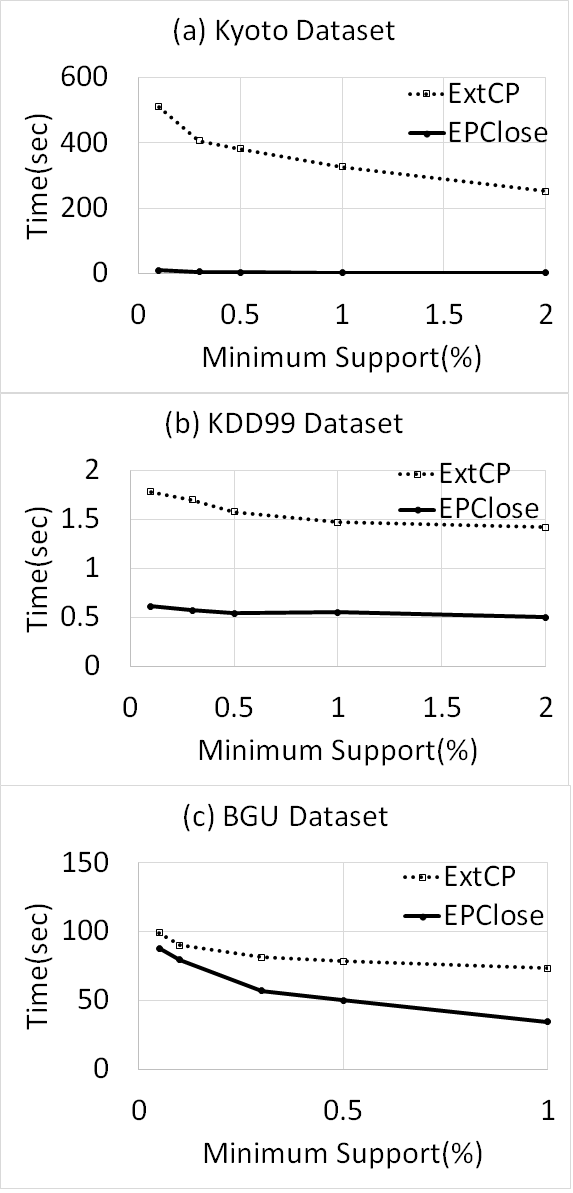}
		\vspace{-1.9em}
		\caption{Run time for different datasets.}
		\Description{Processing Time for different datasets.}
		\label{fig:ProcessingTime}
	\end{minipage}
	\hspace{0.1cm}
	\begin{minipage}[b]{0.47\linewidth}
		\centering
		\includegraphics[height=6.1cm,width=\textwidth]{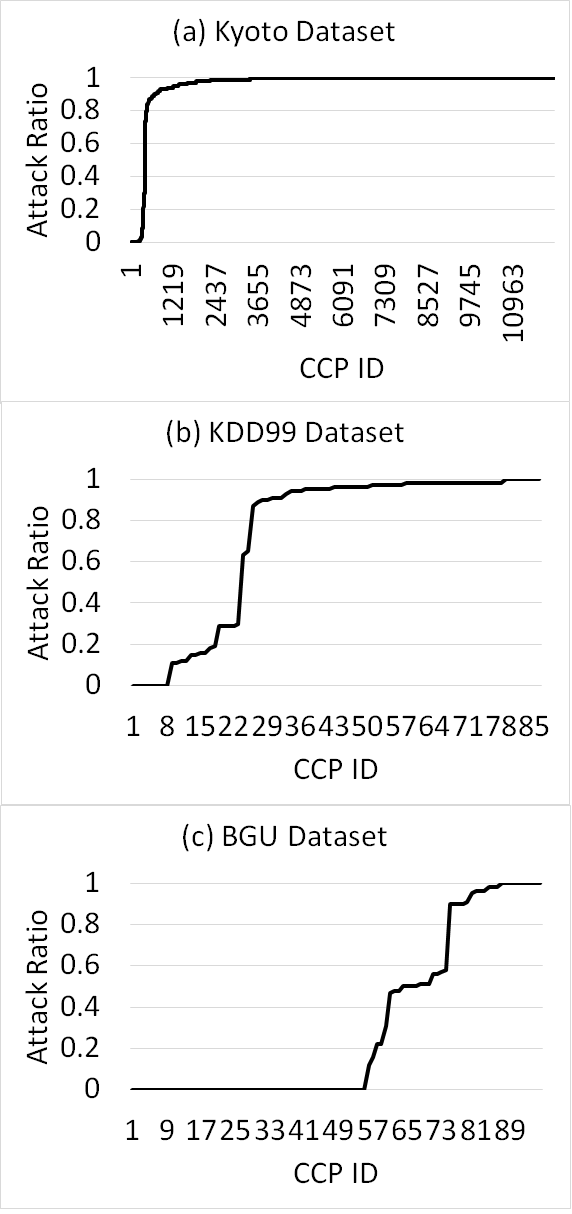}
		\vspace{-1.9em}
		\caption{Attack ratio for different datasets.}
		\Description{Attack ratio for different datasets.}
		\label{fig:AttackRatio}
	\end{minipage}
	
\end{figure}
\vspace{-0.5em}
Figure \ref{fig:ProcessingTime} illustrates the runtime of each algorithm in the three datasets. Although our generated patterns are the same as for \textit{ExtCP}, our algorithm considerably outperforms it, and obtains speed-up rates of up to 100 over \textit{ExtCP}. In Kyoto, with minimum support of 0.1$\%$, the processing time is only 10 seconds for our algorithm, while this time grows to 500 sec for \textit{ExtCP}. The reason why this difference is less for the BGU dataset is that BGU was discretized into two bins, causing many duplicate transactions. As a result, the number of generated CCPs reduces. So the cost of matching in \textit{ExtCP} is comparable to the cost of FP-tree construction.  

In Figure \ref{fig:AttackRatio}, we evaluate the quality of extracted CCPs for network traffic analysis. Although our approach for CCP generation is unsupervised, we use the labels in the target dataset for evaluation. Figure \ref{fig:AttackRatio} shows the \textit{attack ratio} per CCP for three datasets. Attack ratio is \textit{the probability that a CCP belongs to the attack class in the target dataset} \cite{chavary:summarizing}. The derived CCPs are for a minimum support of 0.1\%. It is clear from the graphs that a substantial portion of CCPs are pure patterns. In Kyoto, 97$\%$ of CCPs can uniquely distinguish between classes. This number is 78$\%$ and 80$\%$ for KDD'99 and BGU respectively. It is worth noting that the CCPs for Kyoto (which is a real-life dataset) were almost all pure. 

\section{Conclusion and Future Work}

In this paper, we investigated the suitability of closed patterns for CPM. We proposed a new algorithm, called \textit{EPClose} that uses a revised version of the \textit{FP-tree} data structure to derive all CCPs directly from the closed patterns. Our experimental results show that \textit{EPClose} is much faster than the existing \textit{ExtCP} algorithm. We also show that CCPs have strong discriminative power in detecting pure network traffic patterns. As future work, we will compare the performance of our work with other approaches, and investigate how to efficiently mine CCPs online over data streams.

%\begin{acks}
%To Robert, for the bagels and explaining CMYK and color spaces.
%\end{acks}

\bibliographystyle{ACM-Reference-Format}
\bibliography{refrenceA}
%\bibliography{sample-base}

\balance

%\appendix

\typeout{get arXiv to do 4 passes: Label(s) may have changed. Rerun}
\end{document}